\documentclass[journal=jacsat,
manuscript=article,
]{achemso}
\usepackage{chemformula} 
\usepackage[T1]{fontenc} 
\usepackage{mathrsfs,amsmath,amssymb,bm}
\usepackage{graphicx}
\usepackage{subfigure}
\usepackage[ruled,linesnumbered]{algorithm2e}
\usepackage{multirow}
\usepackage{booktabs}
\usepackage{color}
\usepackage{diagbox}
\usepackage{graphicx}
\usepackage{slashbox}
\usepackage[utf8]{inputenc}
\DeclareUnicodeCharacter{2032}{\textprime}

\SectionsOn

\newcommand{\br}{\bm{r}}
\newcommand{\bu}{\mathbf{u}}

\newcommand{\bR}{\bm{R}}

\newcommand{\bS}{\boldsymbol{S}}

\newcommand{\bM}{\mathbf{M}}
\newcommand{\bI}{\boldsymbol{I}}
\newcommand{\Rmnum}[1]{\expandafter\@slowromancap\romannumeral #1@}

\SectionNumbersOn

\author{Dan Wei}
\affiliation[Xiangtan University]
{Hunan Key Laboratory for Computation and Simulation in Science and Engineering, Key Laboratory of Intelligent Computing and Information Processing of Ministry of Education, School of Mathematics and Computational Science, Xiangtan University, Xiangtan, Hunan, 411105, China.}
\author{Zhijuan He}
\affiliation[Xiangtan University]
{Hunan Key Laboratory for Computation and Simulation in Science and Engineering, Key Laboratory of Intelligent Computing and Information Processing of Ministry of Education, School of Mathematics and Computational Science, Xiangtan University, Xiangtan, Hunan, 411105, China.}
\author{Yunqing Huang}
\affiliation[Xiangtan University]
{Hunan Key Laboratory for Computation and Simulation in Science and Engineering, Key Laboratory of Intelligent Computing and Information Processing of Ministry of Education, School of Mathematics and Computational Science, Xiangtan University, Xiangtan, Hunan, 411105, China.}
\author{An-Chang Shi}
\email{shi@mcmaster.ca}
\affiliation[McMaster UniVersity]
{Department of Physics and Astronomy, McMaster University, Hamilton, Ontario L8S 4M1, Canada.}
\author{Kai Jiang}
\affiliation[Xiangtan University]
{Hunan Key Laboratory for Computation and Simulation in Science and Engineering, Key Laboratory of Intelligent Computing and Information Processing of Ministry of Education, School of Mathematics and Computational Science, Xiangtan University, Xiangtan, Hunan, 411105, China.}
\email{kaijiang@xtu.edu.cn}

\title[An \textsf{achemso} demo]
{Phase Behaviour of X-Shaped Liquid Crystalline Molecules}

\abbreviations{IR,NMR,UV}

\begin{document}



\begin{abstract}
X-shaped liquid crystalline molecules (XLCMs) are obtained by tethering two flexible end A-blocks and two flexible side B-blocks to a rigid backbone (R). A rich array of ordered structures can be formed from XLCMs, driven by the competition between the interactions between the chemically distinct blocks and the molecular connectivity. Here, we report a theoretical study on the phase behaviour of XLCMs with symmetric and asymmetric side blocks by using the self-consistent field theory (SCFT). A large number of ordered structures, including stable smectic-A, triangle-square, pentagon and giant polygon, are obtained as solutions of the SCFT equations. Phase diagrams of XLCMs as a function of the total length and asymmetric ratio of the side chains are constructed. For XLCMs with symmetric side blocks, the theoretically predicted phase transition sequence is in good agreement with experiments. For XLCMs with a fixed total side chain length, transitions between layered structure to polygonal phases, as well as between different polygonal phases, could be induced by varying the asymmetry of the side chains. The free energy density, domain size, side-chain stretching , and molecular orientation are analyzed to elucidate mechanisms stabilizing the different ordered phases. 
\end{abstract}

\section{Introduction}
\label{sec:intrd}
Multiblock liquid crystalline molecules (LCMs) are heterogeneous macromolecules containing rigid and flexible components or blocks. The presence of rigid and flexible blocks enables the simultaneous emergence of liquid crystalline order and microphase separation in the same system, resulting in the formation of various ordered phases. The rich phase behaviour of LCMs make them highly desirable in various applications, ranging from advanced materials to cutting-edge technologies~\cite{woltman2007liquid,larsen2003optical,wang2016stimuli,yang2021beyond}. The performances of liquid crystal materials depend crucially on their microscopic structures. In-depth exploration of the self-assembly structures of LCMs provides a solid foundation for the design and synthesis of new materials.

Among the numerous types of LCMs, X-shaped liquid crystalline molecules (XLCMs), composed of a rigid backbone with two end blocks and two side blocks, have been extensively studied expreimentally~\cite{kieffer2008x-shaped,cheng2009liquid,zeng2011complex,gao2012dithiophene,tschierske2012complex,liu2012arrays,cheng2013transition,gao2015synthesis,werner2015dendritic,lechner2015temperature,achilles2016self,poppe2017formation,poppe2017transition,poppe2017emergence,poppe2017effects,nurnberger2019soft,fall2019an,poppe2020different,poppe2020a,poppe2021emergence,saeed2022rhombic}. It has been observed that the phase behaviour of XLCMs could be regulated by varying the length of the side chains. For XLCMs with symmetric side chains, increasing the total length of the side chains results in several distinct phase transition sequences, including Square $\rightarrow$ Hexagon $\rightarrow$ Lamella phase~\cite{kieffer2008x-shaped}, Triangular $\rightarrow$ Diamond $\rightarrow$ Square phase~\cite{saeed2022rhombic}, and Nematic phase $\rightarrow$ Smectic phase $\rightarrow$ bicontinuous cubic phase $\rightarrow$ Triangular phase~\cite{poppe2017formation,poppe2017transition,poppe2017effects}. For XLCMs with asymmetric side chains, a Square phase emergences by varying the asymmetry of side chains~\cite{kieffer2008x-shaped}. Moreover, for XLCMs with chemically different side chains, various complex tilings have been observed~\cite{zeng2011complex,liu2012arrays,poppe2017effects,nurnberger2019soft,fall2019an}.

Compared with the considerable experimental efforts, theoretical and simulation studies of the phase behaviour of XLCMs have been very limited. Specifically, a dissipative particle dynamics simulation has been used to investigate layered and polygonal phases of XLCMs with chemically identical and different side chains. For identical side chains, the phase sequence of Square $\rightarrow$ Hexagon $\rightarrow$ Lamella has been observed when the side chains length is increased. For chemically different side chains, several multi-color tiling patterns have been observed~\cite{glettner2008liquid,bates2009dissipative,bates2009dissipativemulti-color,bates2011computer}. Furthermore, a theoretical study using the self-consistent field theory (SCFT) has been carried out to explore Archimedean tilings self-assembled from X-shaped molecules with a rigid backbone and two chemically different side chains~\cite{liu2018archimedean}. A systematic exploration of the phase behaviour of XLCMs is lacking and our understanding of the structural stability of XLCMs remains incomplete.

In this study, we fill this gap by carrying out a theoretical study on the phase behaviour of XLCMs composed of three chemically distinct components, {\it i.e.} a rigid backbone (R), two flexible end chains (A) and two flexible side chains (B). It is noted that the XLCMs used in the  previous SCFT study~\cite{liu2018archimedean} possess two chemically different side chains without the end chains. The model XLCMs used in the current study is designed to accurately depict the X-shaped molecules, composed of a rigid core, two terminal glycerol groups and two flexible n-alkyl or semiperfluorinated side chains, used in previous experiments.~\cite{kieffer2008x-shaped,saeed2022rhombic,poppe2017formation} It is well-established that the SCFT is a powerful approach for studying the equilibrium phase behaviour of inhomogeneous macromolecular systems~\cite{morse2002semiflexible,matsen1998liquid,song2009new,song2010phase,gao2011self-assembly,jiang2013influence,zhu2013self-assembly,cai2017liquid,liu2018archimedean,you2023stability}. We have extended the SCFT formalism to XLCMs and derived a set of SCFT equations that should be solved numerically. Due to the complexity of XLCMs, solving the SCFT equations is computationally challenging. To address this challenges, we developed an efficient and accurate parallel algorithm for the problem, enabling us to obtain a set of solutions corresponding to different phases of the system. The relative stability of the different ordered phases is examined by comparing their free energy density. A phase diagram of the XLCMs is constructed as a function of the total length of side chains and the asymmetry ratio. Our findings not only replicate the experimental observation but also uncover several new stable ordered structures. More importantly, we provide a detailed analysis of the stability mechanisms.

\section{Theoretical Framework}

\begin{figure}[!htbp]
  \includegraphics[width=7cm]{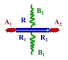}
  \caption{Schematics of an XLCM composed of a rigid backbone block $R$ (blue) connected with two flexible end blocks $A$ (red) and two flexible side blocks $B$ (green).}
  \label{fig:xshape}
\end{figure}

We consider an incompressible melt of $n$ XLCMs with a degree of polymerization $N$ in a volume $V$. Each XLCM (Figure\,\ref{fig:xshape}) consists of six blocks constructed from three chemically distinct monomers ($A,B,R$). The two semiflexible blocks ($R=R_{1}+R_{2}$) form the rigid backbone, which is connected with two flexible end blocks ($A_{1}$ and $A_{2}$) and two flexible side blocks ($B_{1}$ and $B_{2}$). The white arrow along the backbone represents the orientation of block $R$ that is used in the SCFT modeling. The number of segments for the six blocks is denoted by $N_{i} = f_{i}N (i=A_{1}, A_{2}, B_{1}, B_{2}, R_1, R_2)$, where $f_{i}$ is the volume fraction of the $i$th-block. It is noted that $f_{A_{1}} + f_{A_{2}} + f_{B_{1}} + f_{B_{2}} + f_{R_1} + f_{R_2} = 1$ and $N = N_{A_{1}} + N_{A_{2}} + N_{B_{1}} + N_{B_{2}} + N_{R_1} +N_{R_2}$. We use $b_{\alpha}$ to denote the statistical segment length of monomer $\alpha$ ($\alpha=A,B,R$). A space curve $\bR(s)$ represents the chain conformation, where $s$ is a chain trajectory variable scaled by $N$. The unit tangent vector, $\bu(s)=d\bR(s)/ds$, denotes the orientation of the semiflexible chain at contour position $s$. Here, we use Gaussian chain and wormlike chain to model flexible and semiflexible chains, respectively.~\cite{fredrickson2006equilibrium}.  

The central quantity of SCFT is a set of propagators describing the probability distribution of the segments. For flexible chains obeying the Gaussian statistics, the propagators are obtained as solutions of the modified diffusion equations (MDEs),
\begin{equation}
    \begin{split}
        \frac{\partial}{\partial s}q_{\alpha}(\br,s) &= \nabla_{\br}^{2} q_{\alpha}(\br,s) - \omega_{B}(\br) q_{\alpha}(\br,s),\\q_{\alpha}(\br,0)&=1,\quad
        0\le s\le f_{\alpha}, \quad \alpha=B_{1},B_{2},
        \label{eq:B}
    \end{split}
\end{equation}
\begin{equation}
    \begin{split}
        \frac{\partial}{\partial s}q_{\gamma}(\br,s) &= \epsilon^{2}\nabla_{\br}^{2} q_{\gamma}(\br,s) - \omega_{A}(\br) q_{\gamma}(\br,s),\\q_{\gamma}(\br,0)&=1,\quad
        0\le s\le f_{\gamma},\quad \gamma=A_{1},A_{2},
        \label{eq:A}
    \end{split}
\end{equation}
where $\epsilon=b_{A}/b_{B}$ is the conformational asymmetric ratio of monomers $A$ and $B$, $w_{B}(\br)$ and $w_{A}(\br)$ are the mean fields representing the average interactions exerted to the $B-$ and $A-$ monomers, respectively. The propagators $q_{\alpha}(\br,s)$ and $q_{\gamma}(\br,s)$ describe the probability of the $s$-th chain segment at spatial position $\br$ in mean fields $w_{B}(\br)$ and $w_{A}(\br)$, respectively. 
For rigid rod blocks modelled as semiflexible chains, the propagators satisfy the ``convection diffusion'' equations (CDEs),
\begin{equation}
    \begin{split}
        \frac{\partial }{\partial s}q_{R_{1}}(\br,\bu,s) &=-\beta\bu\cdot \nabla_{\br}q_{R_{1}}(\br,\bu,s)\\
        -\Gamma(\br,\bu)&q_{R_{1}}(\br,\bu,s)+\frac{1}{2\lambda}\nabla^{2}_{\bu}q_{R_{1}}(\br,\bu,s),\\
        q_{R_{1}}(\br,\bu,0)&=\frac{q_{A_{1}}(\br,f_{A_{1}})}{2\pi},\quad 0\le s\le f_{R_{1}},
        \label{eq:R1}
    \end{split}
\end{equation}
\begin{equation}
    \begin{split}
        \frac{\partial }{\partial s}q_{R_{2}}(\br,\bu,s)& =\beta\bu\cdot \nabla_{\br}q_{R_{2}}(\br,\bu,s)\\
        -\Gamma(\br,\bu)&q_{R_{2}}(\br,\bu,s)+\frac{1}{2\lambda}\nabla^{2}_{\bu}q_{R_{2}}(\br,\bu,s),\\
         q_{R_{2}}(\br,\bu,0)&=\frac{q_{A_{2}}(\br,f_{A_{2}})}{2\pi},\quad 0\le s\le f_{R_{2}}, 
        \label{eq:R2}
    \end{split}
\end{equation}
where $\Gamma(\br,\bu) = \omega_{R}(\br)-\bM(\br):(\bu\bu-\frac{1}{2}\bI)$ is the field that is a function of $\br$ and $\bu$, $\beta=\sqrt{6N}b_{R}/b_{B}$ is the aspect ratio of semi-flexible chain. The propagator $q_{\alpha}(\br,\bu,s)$ $ (\alpha=R_{1}, R_{2})$ describes the probability of the $s$-th chain segment at position $\br$ with orientation $\bu$ in mean field $w_{R}(\br)$ and orientational field $\bM(\br)$, and the parameter $\lambda$ measures the stiffness of semiflexible chain.

The backward propagators, $q^{\dagger}_{A_{1}}(\br,s)$, $q^{\dagger}_{A_{2}}(\br,s)$, $q^{\dagger}_{B_{1}}(\br,s)$, $q^{\dagger}_{B_{2}}(\br,s)$,  $q^{\dagger}_{R_{1}}(\br,\bu,s)$, and $q^{\dagger}_{R_{2}}(\br,\bu,s)$ are defined similarly, satisfying similar MDEs and CDEs,
\begin{equation}
    \begin{split}
        \frac{\partial}{\partial s}q^{\dagger}_{\gamma}(\br,s) &= \epsilon^{2}\nabla_{\br}^{2} q^{\dagger}_{\gamma}(\br,s) - \omega_{A}(\br) q^{\dagger}_{\gamma}(\br,s),\\
        0&\le s\le f_{\gamma},\quad \gamma=A_{1}, A_{2},
        \label{eq:A1}
    \end{split}
\end{equation}
\begin{equation}
    \begin{split}
        \frac{\partial}{\partial s}q^{\dagger}_{\alpha}(\br,s) &= \nabla_{\br}^{2} q^{\dagger}_{\alpha}(\br,s) - \omega_{B}(\br) q^{\dagger}_{\alpha}(\br,s),\\
        0&\le s\le f_{\alpha},\quad \alpha=B_{1},B_{2},
        \label{eq:B1}
    \end{split}
\end{equation}
\begin{equation}
    \begin{split}
        \frac{\partial }{\partial s}q^{\dagger}_{R_{1}}(\br,\bu,s) &=\beta\bu\cdot \nabla_{\br}q^{\dagger}_{R_{1}}(\br,\bu,s)
        -\Gamma(\br,\bu)q^{\dagger}_{R_{1}}(\br,\bu,s)\\
        &+\frac{1}{2\lambda}\nabla^{2}_{\bu}q^{\dagger}_{R_{1}}(\br,\bu,s),\quad 0\le s\le f_{R_{1}},
        \label{eq:R10}
    \end{split}
\end{equation}
\begin{equation}
    \begin{split}
        \frac{\partial }{\partial s}q^{\dagger}_{R_{2}}(\br,\bu,s) &=-\beta\bu\cdot \nabla_{\br}q^{\dagger}_{R_{2}}(\br,\bu,s)
        -\Gamma(\br,\bu)q^{\dagger}_{R_{2}}(\br,\bu,s)\\
        &+\frac{1}{2\lambda}\nabla^{2}_{\bu}q^{\dagger}_{R_{2}}(\br,\bu,s),\quad 0\le s\le f_{R_{2}},
        \label{eq:R20}
    \end{split}
\end{equation}
with the initial conditions,
\begin{equation*}
    \begin{split}
    &q^{\dagger}_{A_{1}}(\br,0)=\int q_{R_{1}}(\br,\bu,f_{R_{1}})\ d\bu,\\
    &q^{\dagger}_{A_{2}}(\br,0)=\int q_{R_{2}}(\br,\bu,f_{R_{2}})\ d\bu,\\
    &q^{\dagger}_{B_{1}}(\br,0)=q_{B_{2}}(\br,f_{B_{2}})\int q_{R_{1}}(\br,\bu,f_{R_{1}})q_{R_{2}}(\br,\bu,f_{R_{2}})\ d\bu,\\
    &q^{\dagger}_{B_{2}}(\br,0)=q_{B_{1}}(\br,f_{B_{1}})\int q_{R_{1}}(\br,\bu,f_{R_{1}})q_{R_{2}}(\br,\bu,f_{R_{2}})\ d\bu,\\
    &q_{R_{1}}^{\dagger}(\br,\bu,0)=\frac{1}{2\pi}q_{B_{1}}(\br,f_{B_1})q_{B_{2}}(\br,f_{B_2})q_{R_{2}}(\br,\bu,f_{R_{2}}),\\
    &q_{R_{2}}^{\dagger}(\br,\bu,0)=\frac{1}{2\pi}q_{B_{1}}(\br,f_{B_1})q_{B_{2}}(\br,f_{B_2})q_{R_{1}}(\br,\bu,f_{R_{1}}).
    \end{split}
\end{equation*}
By using of these propagators, the single XLCM partition function $Q$, segment density $\phi_{\alpha}(\br)$ $(\alpha=A,B,R$), and orientational order parameter $\bS(\br)$ can be calculated from the following expressions,
\begin{equation*}
    \begin{split}
        Q &= \frac{1}{V}\int q_{B_{1}}(\br,s) q_{B_{1}}^\dagger(\br,s) d\br, \quad 0\le s\le f_{B_{1}},
    \end{split}
\end{equation*}
\begin{equation*}
    \begin{split}
        \phi_A(\br)&=  \frac{1}{Q}\left(\int_0^{f_{A_1}} q_{A_1}(\br, s) q_{A_1}^{\dagger}(\br, s) d s\right.\\
        &+\left.\int_0^{f_{A_2}} q_{A_2}(\br, s) q_{A_2}^{\dagger}(\br, s) d s\right),\\
        \phi_B(\br)&=  \frac{1}{Q}\left(\int_0^{f_{B_1}} q_{B_1}(\br, s) q_{B_1}^{\dagger}(\br, s) d s\right.\\
        &+\left.\int_0^{f_{B_2}} q_{B_2}(\br, s) q_{B_2}^{\dagger}(\br, s) d s\right),\\
        \phi_R(\br)&=  \frac{2 \pi}{Q}\left(\int_0^{f_{R_1}} \int q_{R_1}(\br, \bu, s) q_{R_1}^{\dagger}(\br, \bu, s) d \bu d s\right. \\
        & \left.+\int_0^{f_{R_2}} \int q_{R_2}(\br, \bu, s) q_{R_2}^{\dagger}(\br, \bu, s) d \bu d s\right),\\
        \bS(\br)&=  \frac{2 \pi}{Q}\int \left(\int_0^{f_{R_1}}  q_{R_1}(\br, \bu, s)(\bu\bu-\frac{1}{2} \bI)\right.\\ 
        &\cdot\left.q_{R_1}^{\dagger}(\br, \bu, s) ds +\int_0^{f_{R_2}} q_{R_2}(\br, \bu, s)\right.\\ 
        &\cdot\left.(\bu\bu-\frac{1}{2} \bI)q_{R_2}^{\dagger}(\br, \bu, s) ds\right)d \bu.
    \end{split}
\end{equation*}
For a given solution of the SCFT equations, the mean-field free energy per chain in unity of $k_{B}T$ is given by,~\cite{fredrickson2006equilibrium, jiang2015analytic}
\begin{equation*}
    \begin{split}
        \frac{F}{nk_{B}T} &= \underbrace{\frac{1}{V}\int  \dfrac{1}{4\zeta_1 N}\mu_1^2(\br)+\dfrac{1}{4\zeta_2N}\mu_2^2(\br)-\mu_{+}(\br)d\br}_{F_{inter}/nk_{B}T}\\
        &+\underbrace{\dfrac{1}{2\eta NV}\int M(\br):M(\br)~d\br}_{F_{orien}/nk_{B}T}
        \underbrace{- \ln Q}_{-TS/nk_{B}T},
    \end{split}
\end{equation*}
where $F_{inter}/nk_{B}T$, $F_{orien}/nk_{B}T$, $-TS/nk_{B}T$ are contributions to the free energy from the segment-segment interactions, orientational interactions, and entropy, respectively. $k_{B}$ is the Boltzmann constant, $T$ is the temperature, $\eta$ is the Maier-Saupe parameter which describes the magnitude of the orientation interaction. $\mu_{+}(\br)$ is the ``pressure'' or chemical potential field that ensures the local incompressibility of the system, $\mu_{1}(\br)$, $\mu_{2}(\br)$ are ``exchange'' chemical potentials fields of the system, which are given by, 
\begin{equation*}
    \begin{split}
        w_{\alpha} &= \mu_{+}(\br) - \sigma_{1\alpha}\mu_{1}(\br)-\sigma_{2\alpha}\mu_{2}(\br),\alpha=A,B,R,\\
        \sigma_{1A}&=\frac{1}{3},\sigma_{1R}=-\frac{2}{3},\sigma_{1B}=\frac{1}{3},\\
        \sigma_{2A}&=\frac{1+\alpha}{3},\sigma_{2R}=\frac{1-2\alpha}{3},\sigma_{2B}=\frac{\alpha-2}{3},\\
        \alpha &= \frac{\chi_{AB}+\chi_{AR}-\chi_{BR}}{2\chi_{AB}},\\
        \zeta_{1}&=\frac{4\chi_{AB}\chi_{BR}-(\chi_{AB}-\chi_{AR}+\chi_{BR})^{2}}{4\chi_{AB}},\\
        \zeta_{2}&=\chi_{AB}.
    \end{split}
\end{equation*}
First-order variations of the free energy functional with respect to the fields $\mu_{1}(\br)$, $\mu_{2}(\br)$, $\mu_{+}(\br)$, $\bM(\br)$ lead to the following of SCFT equations,
\begin{equation*}
    \begin{split}
        & \phi_A(\br)+\phi_B(\br)+\phi_R(\br)=1,\\
        & \mu_1(\br) = 2 \zeta_1 N(\sigma_{1A} \phi_A(\br)+\sigma_{1R} \phi_R(\br)+\sigma_{1B} \phi_B(\br)), \\
        &\mu_2(\br)=2 \zeta_2 N(\sigma_{2A} \phi_A(\br)+\sigma_{2R} \phi_R(\br)+\sigma_{2B} \phi_B(\br)),\\
        & \bM(\br)=\eta N\bS(\br).
    \end{split}
\end{equation*}

Numerically solving these SCFT equations requires an iterative procedure~\cite{xu2013strategy,jiang2015self}. The iteration begins by imposing an initial configuration of the $\bM(\br)$, $w_{\alpha}(\br)$ ($\alpha=A,B,R$), and solving the MDEs (Eqs.\,\ref{eq:B}-\ref{eq:A},\,\ref{eq:B1}-\ref{eq:A1}) and CDEs (Eqs.\,\ref{eq:R1}-\ref{eq:R2},\,\ref{eq:R10}-\ref{eq:R20}) to obtain the propagators. Then these propagators are used to calculate $\phi_{\alpha}(\br)$, $\bS(\br)$ and update $\bM(\br)$, $w_{\alpha}(\br)$. This iteration continues until the monomer densities and mean fields are self-consistent, meeting a prescribed numerical accuracy. The standard for terminating SCFT iterations is that the free energy difference between two successive iterations is less than $10^{-8}$, which is sufficient for determining the stability of ordered structures. 

In the current study, we analyze the relative stability of two-dimensional polygonal phases, which are columnar structures that are homogeneous perpendicular to the polygonal plane. For the two-dimensional phases, the computation can be limited to the two-dimensional space and the orientational calculation is carried out on a unit circle. We use fourth-order backward differentiation and fourth-order Runge-Kutta methods to solve the MDEs and CDEs, respectively, and pseudo-spectral method to handle the spatial and orientational variables with periodic boundary conditions~\cite{cochran2006stability,butcher2003numerical,tzeremes2002efficient,rasmussen2002improved,jiang2010spectral}. A hybrid scheme combining conjugate gradient and alternate iteration methods is used to optimize the computational box and to search for the saddle-points corresponding to ordered structures~\cite{jiang2010spectral,jiang2013discovery,liang2015efficient}. Moreover, we implement a parallel version of these algorithms using the C$++$ language and FFTW-MPI package~\cite{frigo2005design} to accelerate the computations. Further details of these algorithms can be found in our previous work~\cite{he2023theory,tan2024high}. The follwoing set of parameters are chosen based on extensive test: the chain contour is divided into 300 points, with spatial discrete grids of $N_{x} \times N_{y} = 81\times 81$ for layered phases and $N_{x} \times N_{y} = 121\times 121$ for two-dimensional polygonal phases, and the number of orientation points $N_{\theta}=16$. This choice is made to ensure sufficient accuracy of solving the SCFT equations without excessive computational cost. 

\section{Results and discussion}
We focus on the influence of side chain lengths on the self-assemble phase behaviour of XLCMs with symmetric and asymmetric side blocks. The lengths of the different sub-chains are specified by the number of segments, $N_{i}$ ($i=A_1$, $A_2$, $B_1$, $B_2$, $R_1$, $R_2$), of the blocks. Furthermore, we use the ratio of the lengths of the B-blocks, $\kappa=N_{B_1}/N_{B_2}$, to specify the asymmetry of the two side chains, where $N_{B_2}$ is the longer length of side chain and $N_{B} = N_{B_1} + N_{B_2}$ is the total length of side chains. Obviously, $\kappa=1$ corresponds to the case of symmetric side chains, while the limit of $\kappa=0$ indicates the case of one side chain. That is, a small $\kappa$ indicates a strong asymmetry between two side chains. There is a large number of parameters for the XLCMs. In what follows we fix some of these parameters based on extensive numerical simulations. Specifically, we set $\epsilon = 1$, $\lambda = 300$, $\beta = 6$, $\eta = 0.36$, $N_{A_{1}} = N_{A_{2}} = 10$, $N_{R} = 44$, $\chi_{AB} = 0.34$, $\chi_{AR} = 0.38$, $\chi_{BR} = 0.32$ to ensure the stability of layered and polygonal phases. 

\subsection{Equilibrium phases}

\begin{figure*}[!htbp]
  \includegraphics[width=13cm]{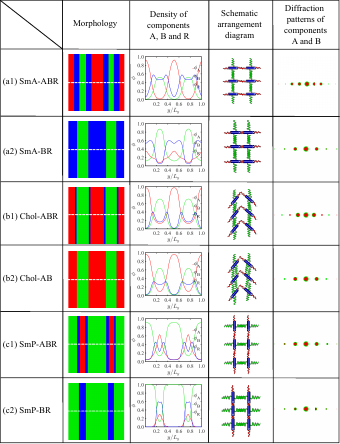}
  \caption{Three types of layered phases in XLCMs, including Smectic-A phases ((a1) SmA-ABR; (a2) SmA-BR)), Cholesteric phases ((b1) Chol-ABR; (b2) Chol-AB), and Smectic-P phases ((c1) SmP-ABR; (c2) SmP-BR). The second column illustrates the morphologies of these phases, where red, green, and blue domains represent the concentrated A, B, and R components, respectively. The density distributions along the white dashed line in the second column are presented in the third column. The fourth column presents schematic arrangements, where the white arrow along the backbone indicates the orientation of block R. The last column exhibits the main diffraction patterns of density distributions A (red) and B (green).}
  \label{fig:lam}
\end{figure*}

\begin{figure*}[!htbp]
  \includegraphics[width=15.2cm]{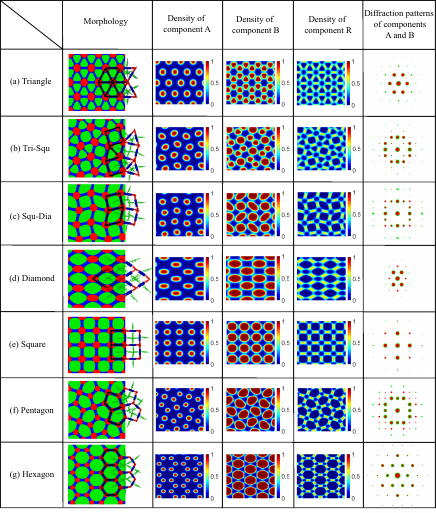}
  \caption{Simple polygonal phases of XLCMs, including (a) Triangle, (b) Tri-Squ (formed by square and triangle in ratio $1 : 2$), (c) Squ-Dia (formed by square and diamond with an internal angle of $60^{\circ}$ in ratio $1 : 1$), (d)  Diamond (internal angle is $60^{\circ}$), (e) Square, (f) Pentagon, and (g) Hexagon. Abbreviations: Tri, Squ, Dia = Triangle, Square, Diamond, respectively. The second column shows their morphologies, where red, green, and blue domains represent the A, B, and R components, respectively. The third to fifth columns exhibit the density distributions of components A, B and R, respectively. The last column displays the diffraction patterns of A (red) and B (green).} 
  \label{fig:simplepolygon}
\end{figure*}

\begin{figure*}[!htbp]
  \includegraphics[width=15.5cm]{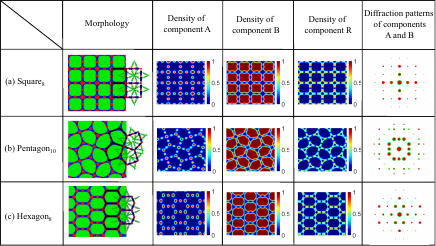}
  \caption{Giant polygonal phases of XLCMs, including (a) Square$_{8}$, (b) Pentagon$_{10}$, and (c) Hexagon$_{8}$. The subfigures presented here have similar meanings as Figure\,\ref{fig:simplepolygon}.} 
  \label{fig:giantpolygon}
\end{figure*}

Obtaining different solutions, corresponding to different ordered phases, of the SCFT equations depends on the library of candidate structures used as the initial configurations. In principle, this library should contain as many candidate phases as possible, which could be derived from experimental observations and numerical simulations~\cite{kieffer2008x-shaped,xu2013strategy,jiang2015self}. In the current study, we construct six layered phases (Figure~\ref{fig:lam}) and ten polygonal phases (Figures~\ref{fig:simplepolygon}, \ref{fig:giantpolygon}) as candidate phases of the XLCMs. 

The layered phases, including Smectic-A (SmA-ABR, SmA-BR), Cholesteric (Chol-ABR, Chol-AB), and Smectic-P (SmP-ABR, SmP-BR), are classified into three categories based on the angle $\theta$ between the normal direction of layer $\bm{n}$ and the orientation of the rigid blocks. The Smectic-A and Semctic-P phases are specified by $\theta = 0$ and $\theta=\pi/2$, respectively. The Cholesteric phase is obtained when the rigid blocks are parallel within each layer but tilted or rotated between layers~\cite{liang2015efficient,he2023theory}. Figure~\ref{fig:lam} gives information about the morphologies, distributions of A, B, and R blocks, schematic arrangements, and diffraction patterns of A (red) and B (green) blocks of these layered phases. The diffraction patterns are calculated from the density distributions of A and B blocks, where the size of dots represents the intensity of diffraction peaks. Here the size of the red dots is reduced by a scaling factor to avoid overlapping with the green dots.

The polygonal phases are divided into simple polygons (Figure~\ref{fig:simplepolygon}) and giant polygons (Figure~\ref{fig:giantpolygon}), based on the number of R-rich domains along side of the polygons. For the simple polygons, their edges are composed of R--rich domains as shown in Figure~\ref{fig:simplepolygon}. Thus the number of polygonal edges is equal to the number of R-rich domains. In this case the rigid R-blocks form the sides of the polygons, connected by the end chains, while the side chains fill the interior of the polygon. For the giant polygons, as could be seen from Figure~\ref{fig:giantpolygon}, their edges are composed of the rigid R-domains and the flexible A-domains. Thus the number of polygonal edges is less than the number of R-rich domains. The naming scheme of these polygonal phases is determined by both their shapes and the number of R-rich domains along their edges (defined as subscripts). For simplicity, the subscripts of simple polygons are omitted, {\it e.g.} the Square$_{4}$ phase is abbreviated as Square. Information about the molecular arrangements is given shown in Figures~\ref{fig:simplepolygon} and \ref{fig:giantpolygon}. It should be noted that the interior of the polygons is B-rich domains filled by the side chains.

\subsection{Phase behaviour of XLCMs with symmetric side chains}
\begin{figure*}[!htbp]
  \includegraphics[width=15cm]{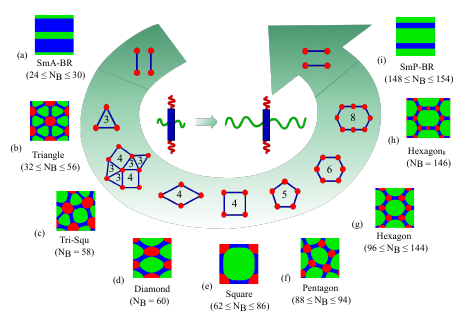}
  \caption{Phase transition sequence when $N_{B}$ is increased. Here $N_{B_{1}}=N_{B_{2}}$, $N_{A_{1}}=N_{A_{2}}=10$, and $N_{R}=44$. (a,i) smectic phases, (b-g) simple polygonal phases, (h) giant polygonal phase. Schematics of the rigid backbone skeleton and domains of A-blocks is shown on the arrow, where the numbers represent the R-rich domains number along polygonal edges. The rigid backbones of SmA-BR are parallel to the normal direction $\bm{n}$ of the layer, while those of SmP-BR are perpendicular to $\bm{n}$.}
  \label{fig:k=1}
\end{figure*}

For XLCMs with symmetric side chains, the lengths of the two side chains are the same, $N_{B_{1}}=N_{B_{2}}$. The total length of the side chains, $N_{B}=N_{B_{1}}+N_{B_{2}}$, is varied to examine their effect on the phase behaviour of the system, focusing on the case with $24 \leq N_{B} \leq 154$. The SCFT results indicate that increasing the length of the side chains results in a series of phase transitions following a generic sequence of smectic-A $\rightarrow$ polygons $\rightarrow$ smectic-P as shown in Figure~\ref{fig:k=1}. Specifically, for short side chains with $24\le N_{B} \le 30$, the layered SmA-BR phase appears because the volume fraction of the B-blocks is too low to form polygonal phases. The structure of the SmA-BR  phase is dominated by the B- and R-domains, whereas the concentration of the A-blocks is peaked at the middle of R-domain as shown in Figure~\ref{fig:lam}(a2). The overall low concentration of the A-blocks is due to their uniform distribution in the R-domain. The layered SmA-BR phase persists till $N_B \leq 30$ and then is taken over by the polygonal phases for larger $N_B$.

Several polygonal structures, from Triangle to Hexagon$_{8}$, appear when the length of the side chains is increased, largely driven by the increased volume of the B-blocks. The increase of the B-rich domain size $S_{B-rich}$, computed using a Monte Carlo integration method as described in reference~\cite{epperson2013an}, is shown in Table~\ref{tab:B-rich}. As $N_B$ increases from $32$ to $146$, $S_{B-rich}$ expands from $0.030$ to $0.172$. Consequently, a larger interior space is required to accommodate these longer side chains and to alleviate their packing frustration~\cite{cao2023understanding}.
This competition leads to the phase transition sequence of Triangle $\rightarrow$ Tri-Squ $\rightarrow$ Diamond $\rightarrow$ Square $\rightarrow$ Pentagon $\rightarrow$ Hexagon $\rightarrow$ Hexagon$_{8}$. When the $N_B$ is increased further to $148$, the side chains are too long to be accommodated in polygons, and leading to the formation of the layered SmP-BR phase.

The theoretically predicted phase transition sequence is in qualitative agreement with the experimental observations of Square $\rightarrow$ Hexagon $\rightarrow$ Lamella\,\cite{kieffer2008x-shaped} and Triangle $\rightarrow$ Diamond $\rightarrow$ Square\,\cite{saeed2022rhombic}. More interestingly, the SCFT results reveal several stable phases, including SmA-BR, Tri-Squ, Pentagon, and Hexagon$_{8}$, which have not been observed in experiments. The discrepancy between theory and experiments may be originated from the choice of system parameters in theory and the narrow stable windows of these phases.

\begin{table*}[!htbp]
  \centering
  \fontsize{12}{19}\selectfont
  \caption{The size of B-rich domain $S_{B-rich}$ from Triangle $\rightarrow$ Tri-Squ $\rightarrow$ Diamond $\rightarrow$ Square $\rightarrow$ Pentagon $\rightarrow$ Hexagon $\rightarrow$ Hexagon$_{8}$, with the increase of $N_{B}$.}
  \label{tab:B-rich}
  \resizebox{1.0\linewidth}{!}{
  \begin{tabular}{|c|c|c|c|c|c|c|c|}
  \hline
	\diagbox[width=4.1em]{\hspace{1em}}{} &Triangle &Tri-Squ &Diamond &Square &Pentagon &Hexagon&Hexagon$_{8}$\\ 
  \hline
    $N_{B}$ &32 - 56 &58 &60 &62 - 86 &88 - 94 &96 - 144 &146\\
  \hline
  $S_{B-rich}$ &0.030 - 0.054 &0.079 &0.114 &0.115 - 0.138 &0.139 - 0.144 &0.145 - 0.170&0.172\\
  \hline
  \end{tabular}}
\end{table*}

\begin{figure}[!htbp]
  \centering
  \includegraphics[width=14.5cm]{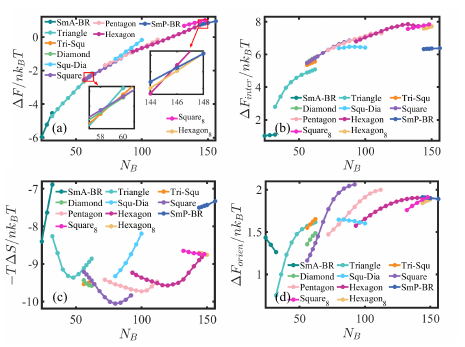}
  \caption{(a) Free energy, (b) monomer interaction energy, (c) entropic contribution energy and (d) orientation interaction energy of various ordered phases relative to free energy of the disordered phase as a function of $N_{B}$ for symmetric side chains fixed $N_{B_1} = N_{B_2}$.}
  \label{fig:energy-k=1}
\end{figure}

The relative stability of the different phase is determined by comparing their free energy. It is interesting to examine the roles played by the various components of the free energy. To this end, we decompose the free energy into three parts: monomer interaction energy $F_{inter}/nk_{B}T$, orientation interaction energy $F_{orien}/nk_{B}T$, and entropic contribution energy $- T S/nk_{B}T$. The free energy and its components for the phases considered in the current study are plotted in Figure~\ref{fig:energy-k=1}, where the free energy of the disordered phase is used as a reference. Compared with the layered phases, the polygonal phases have a larger number of A-, B- and R-rich domains, resulting in a higher monomer interaction energy than that of the layered phases (Figure~\ref{fig:energy-k=1}\,(b)). The A-, B-, and R-rich domains facilitate the flexible arrangement of XLCMs, while hindering the parallel alignment of rigid blocks. This leads to a lower entropic contribtion (Figure~\ref{fig:energy-k=1}\,(c)) and a slightly higher orientation interaction energy (Figure~\ref{fig:energy-k=1}\,(c)) for the polygonal phases. As illustrated in Figure~\ref{fig:energy-k=1}, the phase transition sequence shown in Figure\,\ref{fig:k=1} emerges due to the subtle competition among the monomer interaction energy $F_{inter}/nk_{B}T$, orientation interaction energy $F_{orien}/nk_{B}T$, and entropy contribtuion energy $-T S/nk_{B}T$.

\begin{figure*}[!htbp]
  \centering
  \includegraphics[width=15cm]{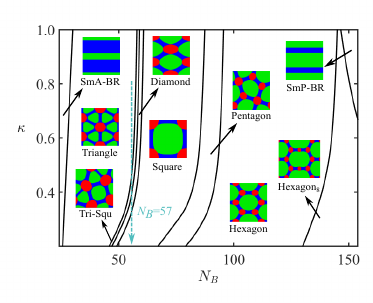}
  \caption{Phase diagram of XLCMs in the $N_{B}$-$\kappa$ plane. The cyan dashed line marks the phase transition sequence of $N_{B}=57$ as the decrease of $\kappa$.}
  \label{fig:phasediagram}
\end{figure*}

\subsection{Phase behaviour of XLCMs with asymmetric side chains}
We now examine the phase behaviour of XLCMs with asymmetric side chains by constructing phase diagram of the system as a function of the total length of the side chains $N_B$ and the asymmetric ratio $\kappa$ (Figure~\ref{fig:phasediagram}). For XLCMs with weak side chain asymmetry ($0.7<\kappa<1$), a phase transition sequence of SmA-BR $\rightarrow$ Triangle $\rightarrow$ Tri-Squ $\rightarrow$ Diamond $\rightarrow$ Square $\rightarrow$ Pentagon $\rightarrow$ Hexagon $\rightarrow$ Hexagon$_8$ $\rightarrow$ SmP-BR is observed. The phase boundaries in this case are insensitive to the value of $\kappa$. On the other hand, the phase boundary curves toward smaller $N_B$ values when the side chain asymmetry of the XLCMs is larger ($\kappa<0.4$). In this case, phase transitions from the layered SmA-BR phase to polygonal phases could be induced by increasing the side chain asymmetry or decreasing $\kappa$, following several sequences depending on $N_B$, {\it i.e.} SmA-BR $\rightarrow$ Triangle ($N_{B}=28$); Triangle $\rightarrow$ Tri-Squ $\rightarrow$ Diamond $\rightarrow$ Square ($51\leq N_{B}\leq 57$); Square $\rightarrow$ Pentagon $\rightarrow$ Hexagon ($N_{B}=87$); Hexagon $\rightarrow$ Hexagon$_8$ ($134\leq N_{B}\leq 144$); SmP-BR $\rightarrow$ Hexagon$_8$ ($148\leq N_{B}\leq 154$).

\begin{figure}[!htbp]
  \centering
  \includegraphics[width=14cm]{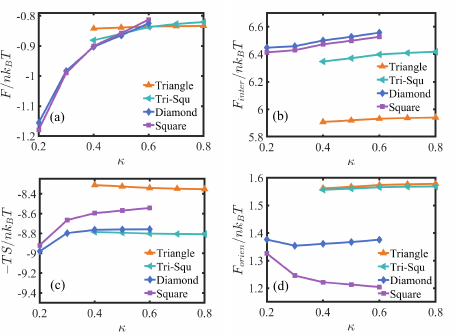}
  \caption{(a) Free energy, (b) monomer interaction energy, (c) entropic contribution energy and (d) orientation interaction energy of Triangle, Tri-Squ, Diamond and Square along changing $\kappa$, fixed $N_{B}=57$.}
  \label{fig:NB=57}
\end{figure}

\begin{figure*}[!htbp]
  \centering
  \includegraphics[width=15cm]{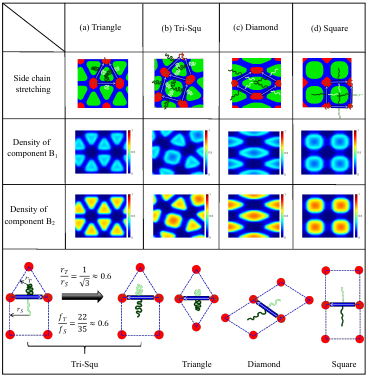}
  \caption{Density of the side $B_{1}$ and $B_{2}$ blocks for different phase and schematics of side chain stretching. (a) Triangle, (b) Tri-Squ, (c) Diamond and (d) Square with $\kappa=0.6$ and $N_B=57$. The dark green and light green side chains represent long $B_{2}$-chain and short $B_{1}$-chain, respectively. The white arrows indicate the orientation of rigid block.}
  \label{fig:k=0.6}
\end{figure*}

To analyze the relative stability of the phases when varying $\kappa$, we present a analysis based on the free energy and the stretching of the side chains for the representative case of $N_{B} = 57$. The phase transition sequence for this case is Triangle $\rightarrow$ Tri-Squ$ \rightarrow$ Diamond $\rightarrow$ Square, as shown by the cyan dashed line in Figure~\ref{fig:phasediagram}. The free energy curves of these phases are presented in Figure~\ref{fig:NB=57}. Based on the orientation of the rigid blocks (represented by white arrows), densities of component $B_{1}$ and $B_{2}$, as displayed in the third and fourth rows of Figure~\ref{fig:k=0.6}, respectively, showing the arrangements of the short and long chains. The second row of Figure~\ref{fig:k=0.6} plots the schematics of molecular arrangements for Triangle, Tri-Squ, Diamond, and Square under $\kappa=0.6$. By observation of these molecular arrangements, we find that the interior of Triangle is filled with a larger number of long side chains than short ones. The triangular tile interior of Tri-Squ is filled with more short side chains than long ones, and the quadrilateral tile interior of Tri-Squ is filled with equal number of long and short side chains. The Diamond and Square have the same number of long and short side chains. Furthermore, the ratio of the short chain to the long chain ($N_{B_{1}}/N_{B_{2}}=22/35$) is approximately equal to the ratio of the distance from the edge to the center of the triangle to the distance from the edge to the center of the square ($1/\sqrt{3}$)~\cite{duan2018stability}. In this case, the side chain arrangements of Tri-Squ are more flexible, thus its entropic contribution is lower. On the other hand, the side chains of the Triangle generate packing frustration, while those of the Diamond and Square experience varying degrees of stretching, leading to an increase in their entropic contribution (Figure~\ref{fig:k=0.6}, Figure~\ref{fig:NB=57}\,(c)).

\begin{figure*}[!htbp]
  \centering
  \includegraphics[width=16.5cm]{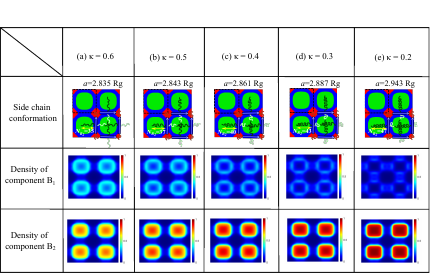}
  \caption{Density of the side $B_{1}$ and $B_{2}$ blocks of Square for a fixed $N_{B}=57$ and different values of $\kappa$ and schematics of side chain stretching. The edge length $a$ for different $\kappa$ is given in the second raw. The dark green and light green side chains represent long $B_{2}$-chain and short $B_{1}$-chain, respectively. The white arrows indicate the orientation of rigid block. The yellow double arrows indicate the stretching the long side chains.}
  \label{fig:k-square}
\end{figure*} 

When the lengths of the two side chains are different, their conformation and degree of stretching are different, which in turn affecting the formation of polygonal phases. These properties are analyzed by using the Square phase as an example. Based on the orientation of the rigid blocks and densities of the $B_{1}$ and $B_{2}$ blocks shown in the third and fourth rows of Figure~\ref{fig:k-square}, it is seen that the interior of the Square is occupied by the long and short side chains. The $B_1$-concentration is consistently lower than the $B_2$-concentration because $B_2$ is the longer block. It is interesting to observe that the size of the square $a$ increases as $\kappa$ is decreased, demonstrating that the size of the domain is largely determined by the longer block with a length of $N_{B_{2}}=1/(1+\kappa)N_B$. The expansion of the polygonal size when $\kappa$ is decreased is generic for all the polygonal phases. The density profiles shown in Figure~\ref{fig:k-square} also indicate that the shorter side chains are mostly concentrated at the R/B interfaces, whereas the longer side chains are stretched to fill the B-domain. 

\section{Conclusion}
In summary, we formulate a self-consistent field theory for the mutiblock X-shaped liquid crystalline molecules that is composed of a rigid backbone with two end chains and two side chains. The corresponding SCFT equations are solved to obtain various layered and polygonal phases of the system. A comparison of the free energy of the ordered phases is used to construct phase diagrams. For XLCMs with symmetric side chains, the theoretically predicted phase transition sequence is in good agreement with experiments. Furthermore, the theoretical results reveal that several new ordered phases, including the SmA-BP, Tri-Squ, Pentagon and Hexagon$_8$ phases, could become equilibrium phases of the system. For XLCMs with asymmetric side chains, a theoretical phase diagram is constructed, demonstrating the role played by the length ratio of the side chains on the phase behaviour of the system. In particular, varying this ratio could induce order-to-order phase transitions. The stability mechanism is analyzed by examining the free energy, the size of B-rich domain, stretching of side chains, and molecular orientation arrangements of ordered phases. These theoretical findings fill the gap between experimental observation and theoretical study of phase and phase transitions in XLCMs. Furthermore, the results provide a strong validation on the capability of SCFT for the study of molecular systems containing rigid and flexible blocks, thus paving the way to extensive study of this class of fascinating systems.


\section*{Acknowledgement}
This work is supported by the National Key R\& D Program of China (2023YFA1008802), the National Natural Science Foundation of China (12171412), the Science and Technology Innovation Program of Hunan Province (2024RC1052), the Innovative Research Group Project of Natural Science Foundation of Hunan Province of China (2024JJ1008), the Natural Sciences and Engineering Research Council (NSERC) of Canada, and the Postgraduate Scientific Research Innovation Project of Hunan Province (CX20230603).
We are also grateful to the High Performance Computing Platform of Xiangtan University for partial support of this work.

\bibliography{Xshape}

\end{document}